\documentclass{nato}  % Specifies the document style.

\usepackage{txfonts}
\usepackage{sprrefn} % (only with nato.bst)

\usepackage[draft]{hyperref}

\newdisplay{guess}{Conjecture}

\begin{document}

% if you don't want to start at page 9:
\firstpage{1}

\begin{opening}
\title{Electron-electron interaction in carbon nanostructures}
\author{A.I. Romanenko\email{air@che.nsk.su}, O.B. Anikeeva, T.I. Buryakov, E. N. Tkachev, A.V. Okotrub}
\runningauthor{A.I. Romanenko, O.B. Anikeeva, T.I. Buryakov...}
\runningtitle{Electron-electron interaction in carbon
nanostructures} \institute{Nikolaev Institute of Inorganic
Chemistry, Lavrentieva 3, Novosibirsk, 630090 Russia; Novosibirsk State University, Lavrentieva 14, Novosibirsk, 630090 Russia}
\author{V.L. Kuznetsov, A.N. Usoltseva}
\institute{Boreskov Catalysis, Lavrentieva 5, Novosibirsk, 630090
Russia}
\author{A.S. Kotosonov}
\institute{Institute of Carbon, Moscow, Russia}

\begin{abstract}
The electron-electron interaction in carbon nanostructures was studied. A new method which allows to determine the electron-electron interaction constant $\lambda_c$\index{electron-electron interaction} from the analysis of quantum correction to the magnetic susceptibility and the magnetoresistance\index{magnetoresistance} was developed. Three types of carbon materials: arc-produced multiwalled carbon nanotubes (arc-MWNTs), CVD-produced catalytic multiwalled carbon nanotubes (c-MWNTs)\index{carbon nanotubes} and pyrolytic carbon were used for investigation. We found that $\lambda_c$=0.2 for arc-MWNTs (before and after bromination treatment\index{bromination}); $\lambda_c$ = 0.1 for pyrolytic graphite; $\lambda_c >$ 0 for c-MWNTs. We conclude that the curvature of graphene layers in carbon nanostructures leads to the increase of the electron-electron interaction constant $\lambda_c$.\end{abstract}


\end{opening}

\section{Introduction}

The carbon nanostructures are formed of graphene layers which
always have some curvature. As a result, these materials are
characterized by new properties which are not present in graphite
consists of plane graphene layers. The curvature of the graphene
layers influences the electron-electron interaction in these
systems. The most interesting consequence of the graphene layers
curvature is the existence of a superconducting state in bundles
of single-walled carbon nanotubes with diameters of $10\,${\AA} at
temperatures below $T_c \approx$ 1 K \cite{kociak} as well as the
onset of superconductivity at $T_c\approx$ 16 K in nanotubes with
diameters of $4\,${\AA}  \cite{tang} and at $T_c\approx$ 12 K in
entirely end-bonded multiwalled carbon nanotubes \cite{takesue}.
In contrast, in graphite no superconducting state is observed. Y.
Kopelevich {\it et al.} \cite{Kopelevich} proposed that
superconductivity may appear in ideal graphite and that the
absence of superconductivity in real samples is related to the
defects always present in graphite. According to theoretical
predictions of \cite{Gonzalez} topological disorder can lead to an
increase in the density of states at the Fermi surface and to an
instability of an electronic subsystem. These changes in the
electronic system could lead to a superconducting state. However,
such topological disorder leads to a curvature of initially flat
graphene layers. We assume, therefore, that in carbon
nanostructures the superconducting state is related to the
curvature of graphene layers. The curvature of surfaces is always
present in the crystal structure of nanocrystallites. As a result,
in such structures the electron-electron interaction should be
modified. This paper is devoted to the analysis of experimental
data which allows to determine the electron-electron interaction
constant $\lambda_c$ in carbon nanostructures formed by curved
graphene layers.

\section{Experimental methods}
The method of our investigation of the electron-electron
interaction constant is based on the joint analysis of quantum
corrections to the electrical conductance, magnetoconductance and
magnetic susceptibility. For all nanostructures formed by graphene
layers the presence of structural defects leads to the diffusive
motion of charge carriers. As a result, at low temperatures,
quantum corrections to the electronic kinetic and thermodynamic
quantities are observed.  For the one-particle processes (weak
localization - WL \cite{Kawabata, Lee}) these corrections arise
from an interference of electron wave functions propagating along
closed trajectories in opposite directions, provided the lengths
$l$ of these trajectories are less then the phase coherence
lengths $L_\varphi(T) = (D\tau_{\varphi} )^{1/2}$ ($D$ is
the diffusion constant and $\tau _\varphi=T^{-p}$ is the characteristic
time for the loss of phase coherence with an exponent
$p=1\div2$). As a result, the total
conductance of the system is decreased. $L_\varphi(T)$ increases
with decreasing temperature which, in turn, leads to the decrease
of the total conductance. In a magnetic field there is an
additional contribution to the electronic phase, which has an
opposite sign for opposite directions of propagation along the
closed trajectory. As a result, the phase coherence length is
suppressed: $L_B = (\hbar c/2eB)^{1/2} < L_\varphi$. 
Here $L_B = (\hbar c/2eB)^{1/2}$ the magnetic length,  
$c$ is the light velocity, ${\it e}$ - the electron
charge, $B$ - the magnetic field. This leads to
negative magnetoresistance, i.e. to an increase of conductance in
a magnetic field. Quantum corrections also arise from the
interaction between electrons (interaction effects - IE
\cite{Altshuler}). These corrections arise due to the phase memory
between two consecutive events of electron-electron scattering. If
the second scattering event happens at a distance shorter than the
coherence length, $L_{IE} = (D\hbar/k_BT)^{1/2}$, from the first one
($L_{IE}$ being the length on which the information about the changes
of the electronic phases due to the first scattering event is not
yet lost), the second scattering will depend on the first one. As
a result the effective density of states on the Fermi surface
$v_F$ is renormalized. Interaction effects contribute not only to
electrical conductance, but also to thermodynamic quantities
depending on $v_F$ - magnetic susceptibility $\chi$ and heat
capacity $C$.

\section{Results and discussion}
\subsection{Arc-produced multiwalled carbon nanotubes}

A characteristic peculiarity of our arc-MWNTs \cite{Okotrub,Romanenko} is the
preferential orientation of the bundles of nanotubes in the 
plane perpendicular to the electrical arc axis.
The volume samples of our arc-MWNTs show anisotropy in their
electrical conductivity $\sigma_{II}/\sigma_{\bot}\approx 100$
\cite{Okotrub,Romanenko}. $\sigma_{II}$ is the electrical
conductivity in the plane of preferential orientation of the
bundles of nanotubes, $\sigma_{\bot}$ is the conductance
perpendicular to this plane. The average diameter of individual
nanotubes is $d_{MWNT}\approx 140\,${\AA}. According to the 
electron paramagnetic resonance data, the concentration of
paramagnetic impurities in our samples is less than $10^{-6}$.
This excludes a substantial contribution of the impurities to the
susceptibility. The MWNTs brominated at room temperature in
bromine vapour \cite{Romanenko} have a composition of
CBr$_{0.06}$.  The addition of bromine leads to an increase of the
conductivity, which can be attributed to an increase in the
concentration of hole current carriers.

According to experimental and theoretical data, the basic
contribution in $ \chi $ of quasi-two-dimensional graphite (QTDG),
including MWNTs, gives orbital magnetic susceptibility $ \chi _
{or} $  connected with extrinsic carriers (EC).
Figures 1(a) and 2(a) present the magnetic susceptibility $\chi $ of
arc-MWNTs samples before bromination and after bromination as 
a function of temperature respectively. 
Available models well reproduce the temperature dependence of
magnetics susceptibility for MWNTs only at T $>$ 50 K. In the low-temperature region
the experimental data deviate from the theoretical ones. According to
theoretical consideration the magnetic susceptibility $\chi $ of
quasi-two-dimensional graphite (QTDG) is generally contributed by
the component $\chi_D $ \cite{Kotosonov}

\begin{equation}
\chi _{or}(T) = -\frac{5.45\times 10^{-3}\gamma_0^2}{(T+\delta )[2+exp(\eta )+exp(-\eta )]} ,
\end{equation}

where $\gamma_0$ is the band parameter for two-dimensional case,
$\delta $ is the additional temperature formally taking into
account "smearing" the density of states due to electron
nonthermal scattering by structure defects, $\eta $ =
$E_F/k_B(T+\delta )$ represents reduced Fermi level ($E_F$), $k_B$
is the Boltzmann constant. Using an electrical neutrality equation
in the 2D graphite model \cite{Kotosonov} $\eta $ can be derived by
$\eta $ = sgn($\eta _0$)[0.006$\eta _0^4$ - 0.0958$\eta _0^3$ +
0.532$\eta _0^2$ - 0.08$\eta_0$] with an accuracy
no less then $1\%$. The $\eta_0$ is determined by $\eta_0 $ =
$T_0/(T+\delta )$, where $T_0$ being degeneracy temperature of
extrinsic carriers (EC) depends on its concentration $n_0$ only.
The value of $\delta $ can be estimated independently
as $\delta $ = $\hbar $/$\pi k_B\tau _0$ ,
 where $\hbar $ is the Planck constant, $\tau _0$  is a relaxation time of the carrier
nonthermally scattered by defects. Generally, the
number of EC in QTDG is equal to that of scattering centers and
$\delta$ depends only on $T_0$, i.e. $\delta  = T_0/r$, where $r$
is determined by scattering efficiency. These
parameters were chosen to give the best fit of the experimental
data. 

According to theoretical
calculations \cite{Altshuler}, the correction $\Delta\chi_{or}$ to
the orbital susceptibility $\chi_{or}$ in the Cooper channel
dominates the quantum correction to the magnetic susceptibility
$\chi(T,B)$ in magnetic fields smaller than $B_c = (\pi
k_BT/g\mu_B$) ($B_c$ = 9.8 T at 4.2 K). These corrections are
determined by the value and the sign of the electron-electron
interaction constant $\lambda_c$ and are proportional to the
diamagnetic susceptibility of electrons $\chi_{or}$. In graphite
and MWNTs the diamagnetic susceptibility is greater than in any
other diamagnetic material (excluding the superconductors), and
the correction to $\chi_{or}$ should also be large. $\Delta\chi(T) _{or}
= \chi(T)_{or}^{exp} - \chi(T)_{or}$ was found by \cite{Lee,
Altshuler}:

\begin{equation}
 \frac{\Delta \chi_{or} (T)}{\chi _{or}(T)}=
-\frac{\frac{4}{3}(\frac{l_{el}}{h})ln[ln(\frac{T_c}{T})]}
{ln(\frac{k_B T_c\tau _{el}}{\hbar })},(d = 2),
\end{equation}
\begin{equation}
 \frac{\Delta \chi_{or} (T)}{\chi _{or}(T)}=
-\frac{2(\frac{\pi }{6})\xi (\frac{1}{2})
(\frac{k_B T\tau _{el}}{\hbar })^{1/2}}{ln(\frac{T_c}{T})},(d = 3),
\end{equation}

where $\chi(T)_{or}^{exp}$ are the experimental data, $\chi(T)_{or}$
is the result of an approximation of the experimental data in an
interval of temperatures 50 - 400 K by the theoretically predicted
dependence (1) for quasi-two-dimensional graphite \cite{Kotosonov};
value of $\xi (\frac{1}{2})$ $\sim $ 1, $l_{el}$ is the electron
mean free path; $\tau _{el}$ represents the elastic relaxation
time which is about 10$^{-13}$ sec for MWNT ~\cite{Baxendale}; $h
$ is the thickness of graphene layers packet; $d $ denotes the
dimensionality of the system; $T_c$ = $\theta _Dexp(\lambda
_c^{-1}$), where $\theta _D$ is the Debye temperature, $\lambda
_c$ is the constant which describes the electron-electron
interaction in the Cooper channel ($\lambda _c > 0 $ in a case of
electron repulsion). The dependence in Eq. (2) is determined by
$ln[ln(\frac{T_c}{T})]$ term because, at low temperatures, in the
disordered systems, $\tau _{el}$ is temperature independent while
all other terms are constants. The dependence in Eq. (3) is
governed by $\ T^{1/2}$ term as $T_c \gg T $ and, therefore,
$ln(\frac{T_c}{T})$ can be considered as a constant relative to
$T^{1/2}$. The temperature dependence of the magnetic
susceptibility $\chi(T)$ is shown in figure 1 for arc-MWNTs before bromination, 
in figure 2 for arc-MWNTs after bromination, and in figure 3 for crystal graphite. Below 50 K
the deviation of experimental data from the theoretical curve is
observed \cite{Romanenko2, Romanenko3}. The additional
contribution to $\chi(T)$ is presented in Fig. 1(b), 2(b), 3(b); and Fig. 1(c), 2(c), 3(c)
as a function of $ln[ln(\frac{T_c}{T})]$ and $T^{1/2}$
respectively. The $\Delta \chi_{or} (T)/\chi _{or}(T)$ clearly
shows the dependence given by Eq. (2) at low magnetic field and
one given by Eq. (3) at high magnetic field, while at $B$ = 0.5 T
the temperature dependence of $\triangle \chi_{or} (T)/\chi
_{or}(T)$ differs from those two limits. As seen from Fig. 1, at
all magnetic fields applied to the arc-MWNTs before bromination, the absolute
value of $\Delta \chi_{or} (T)/\chi_{or}(T)$  increases with
decreasing temperature as has been predicted for IE in the systems
characterized by electron-electron repulsion \cite{Lee,Altshuler}.
Hence, at $B$ = 5.5 T a crossover from the two-dimensional IE
correction to the three-dimensional one takes place. At lower
magnetic field the interaction length $L_{IE}(T)$ is much shorter 
than the magnetic length $L_B$, which in turn becomes dominant at high
field. An estimation of the characteristic lengths gives
respectively the value of $L_{IE}(4.2K)$ = 130 {\AA } (taking into
account that the diffusion constant $D$ = 1 cm$^2$/s
\cite{Baxendale}) and the value of $L_B$ = 100 {\AA } at $B$ = 5.5
T.

\begin{figure}
\footnotesize
\centerline{\begin{tabular}{@{}cc@{}}
\includegraphics[height=1.7in]{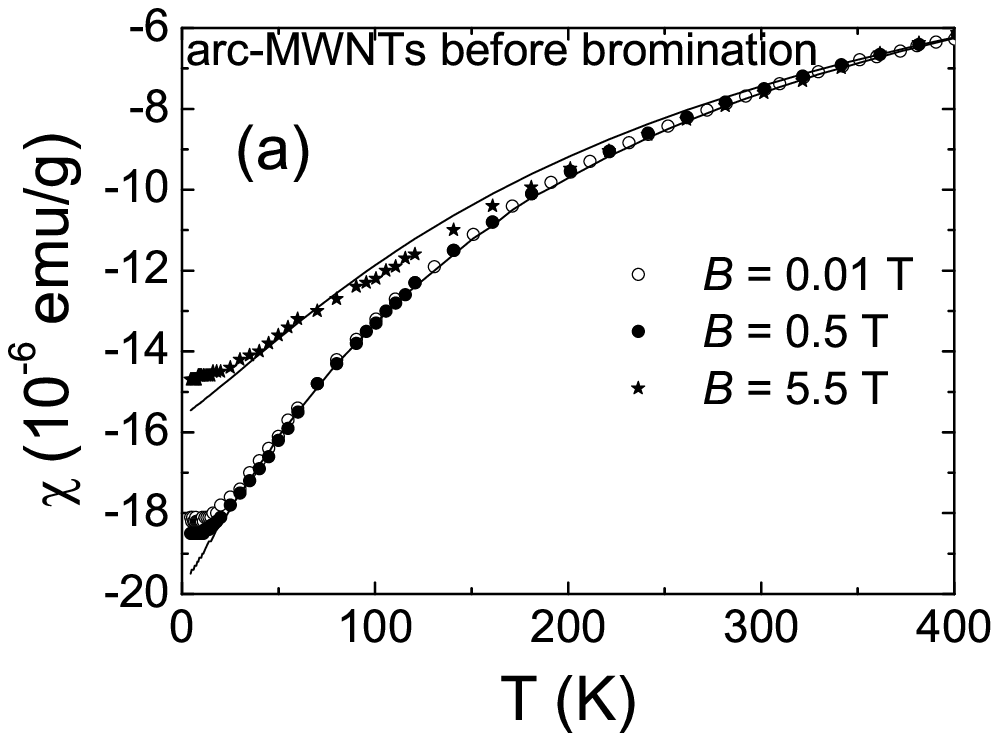}&
\includegraphics[height=1.7in]{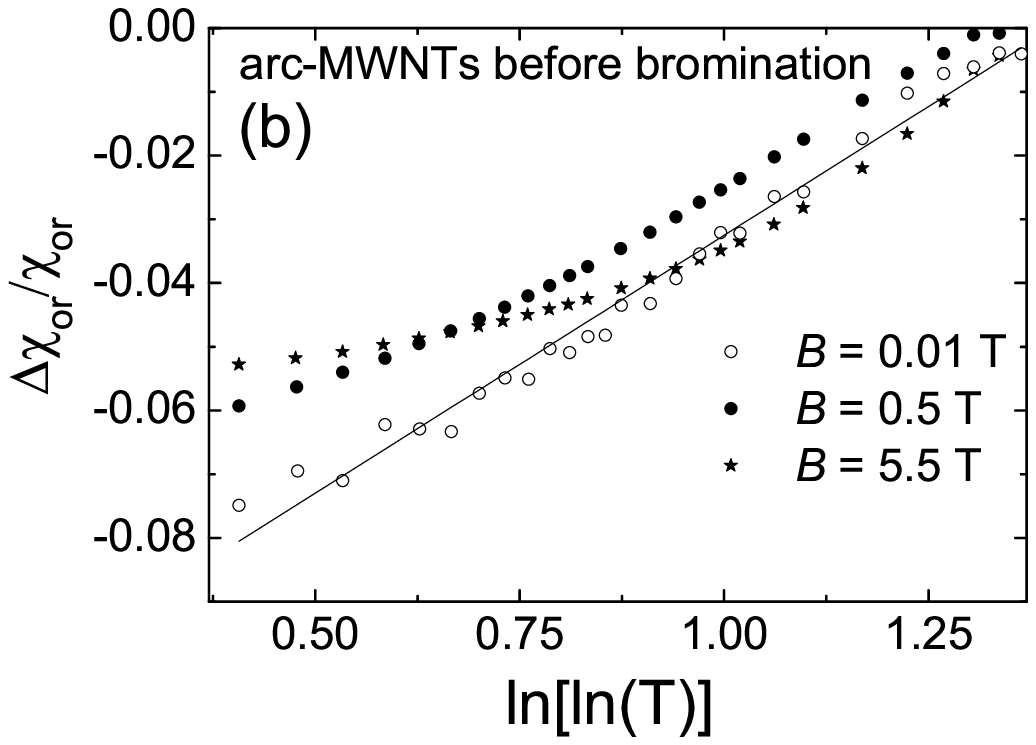}\cr
\includegraphics[height=1.7in]{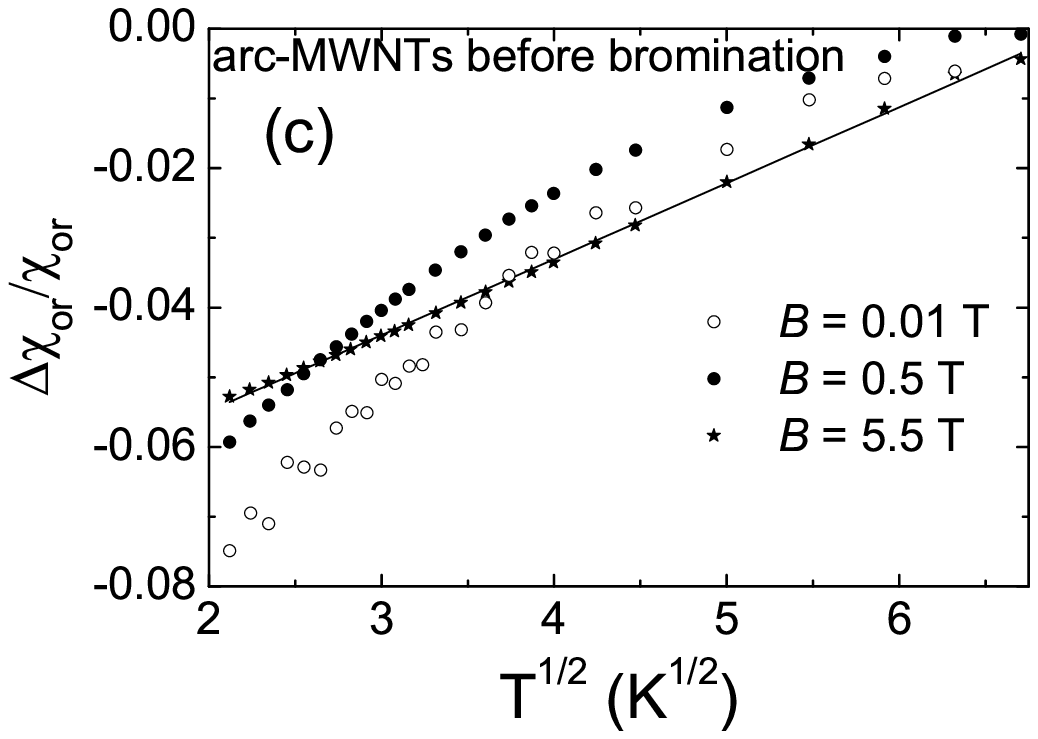}&
\end{tabular}}
\caption{The temperature dependence of magnetic susceptibility
$\chi (T)$ ({\bf a}) and $\Delta \chi_{or} (T)/\chi _{or}(T)$ =
[$\chi (T) -\chi _{or}(T)]/\chi _{or}(T)$ [({\bf b}) and ({\bf
c})] for arc-produced MWNTs sample before bromination. The solid lines are fits: for ({\bf a})
by Eq. (1) in interval 50 - 400 K with parameters; for curve
($\circ$) , $\gamma _0$ = 1.6 eV, $T_0$ = 215 K, $\delta$  = 159
K; for ($\bullet$)  , $\gamma _0$ = 1.6 eV, $T_0$ = 215 K,
$\delta$  = 159 K; for ($\star$)  , $\gamma _0$ = 1.7 eV, $T_0$ =
327 K, $\delta$  = 210 K; by Eq. (2) and Eq. (3) for ({\bf b}) and
({\bf c}) respectively in interval 4.5 - 45 K with parameters
$T_c$ = 10000 K , $l_{el}/a$ = 0.15.} \label{ret}
\end{figure}

A similar dependence of $\Delta \chi_{or} (T)/\chi _{or}(T)$ was
observed for arc-MWNTs after bromination (Fig. 2). 

\begin{figure}
\footnotesize
\centerline{\begin{tabular}{@{}cc@{}}
\includegraphics[height=1.7in]{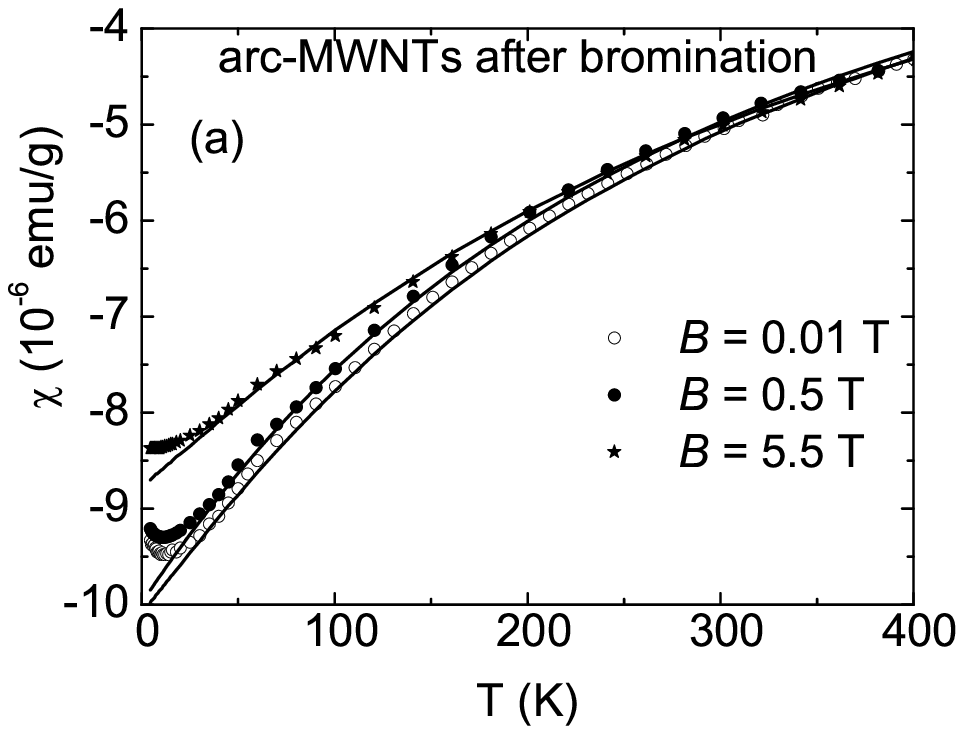}&
\includegraphics[height=1.7in]{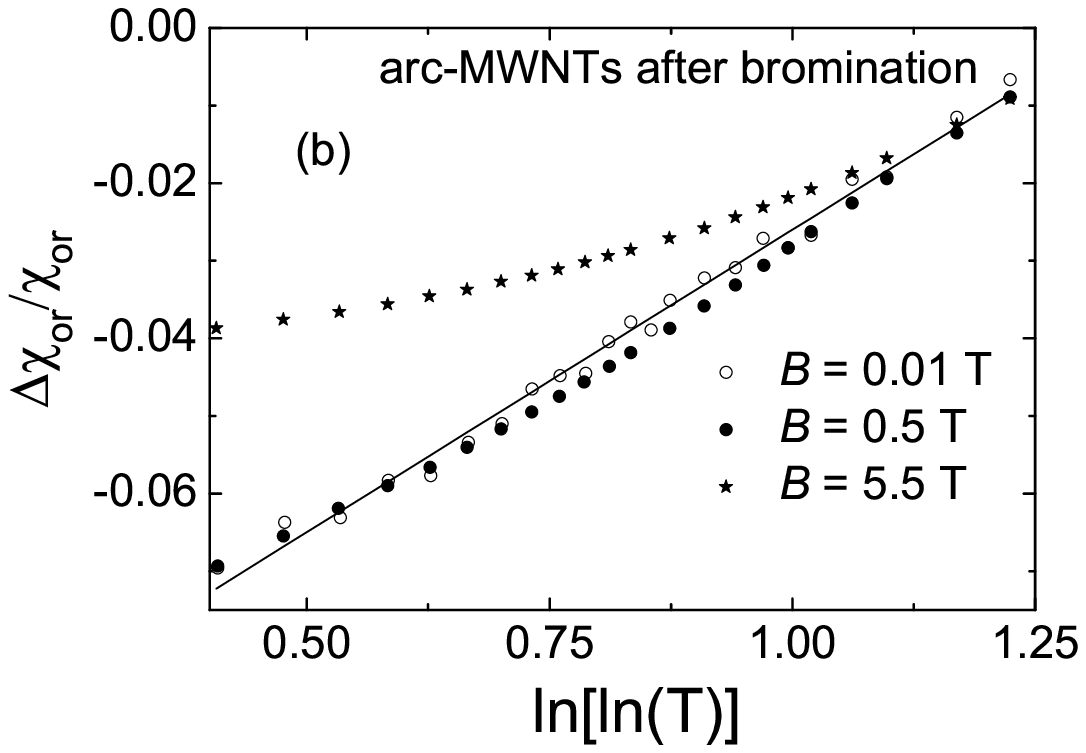}\cr
\includegraphics[height=1.7in]{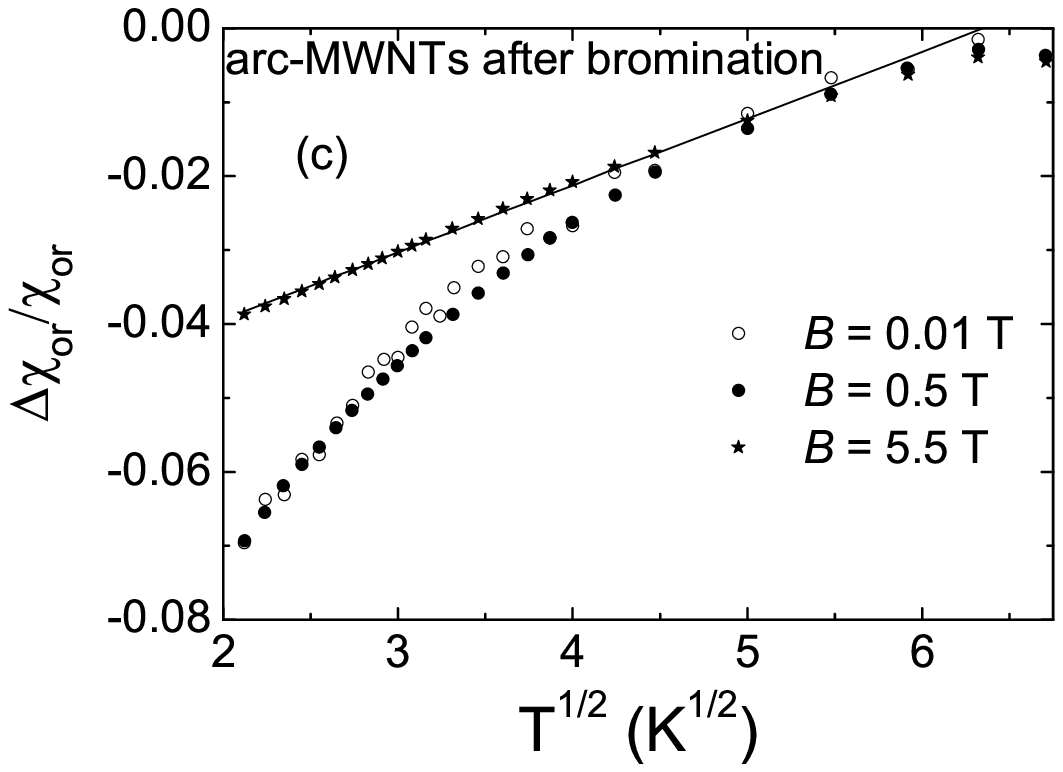}&
\end{tabular}}
\caption{The temperature dependence of magnetic susceptibility
$\chi (T)$ ({\bf a}) and $\Delta \chi_{or} (T)/\chi _{or}(T)$ =
[$\chi (T) -\chi _{or}(T)]/\chi _{or}(T)$ [({\bf b}) and ({\bf
c})] for arc-produced MWNTs sample after bromination. The solid lines are fits: for ({\bf
a}) by Eq. (1) in interval 50 - 400 K with parameters; for curve
($\circ$) , $\gamma _0$ = 1.4 eV, $T_0$ = 340 K, $\delta$  = 252
K; for ($\bullet$)  , $\gamma _0$ = 1.4 eV, $T_0$ = 300 K,
$\delta$  = 273 K; for ($\star$)  , $\gamma _0$ = 1.5 eV, $T_0$ =
435 K, $\delta$  = 325 K; by Eq. (2) and Eq. (3) for ({\bf b}) and
({\bf c}) respectively in interval 4.5 - 45 K with parameters
$T_c$ = 10000 K , $l_{el}/a$ = 0.15.} \label{ret}
\end{figure}

\begin{figure}
\footnotesize
\centerline{\begin{tabular}{@{}cc@{}}
\includegraphics[height=1.7in]{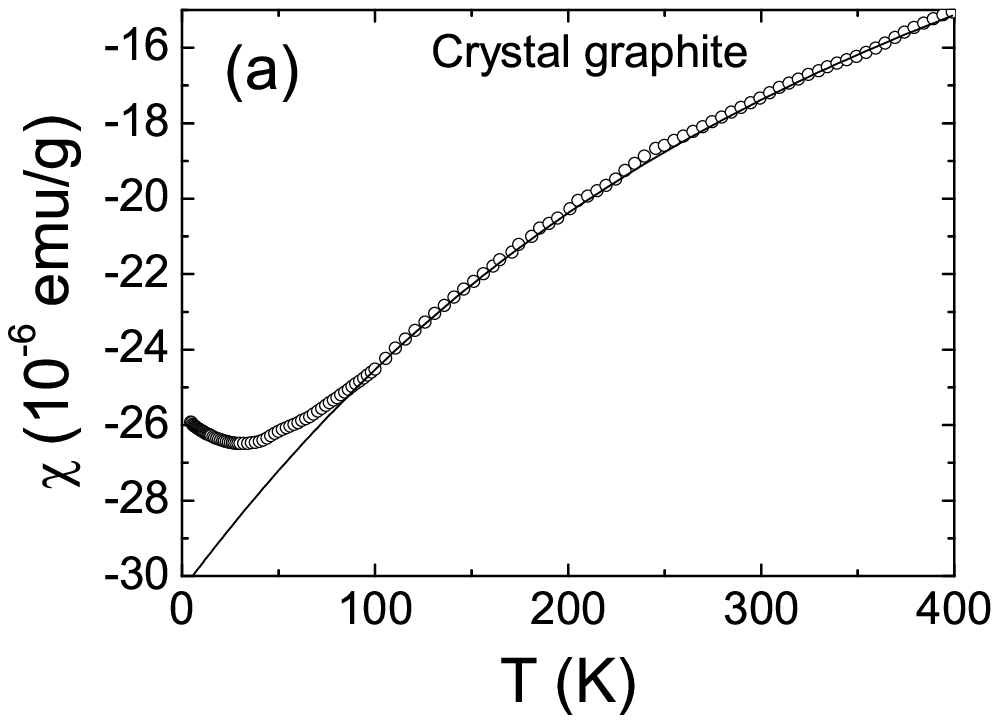}&
\includegraphics[height=1.7in]{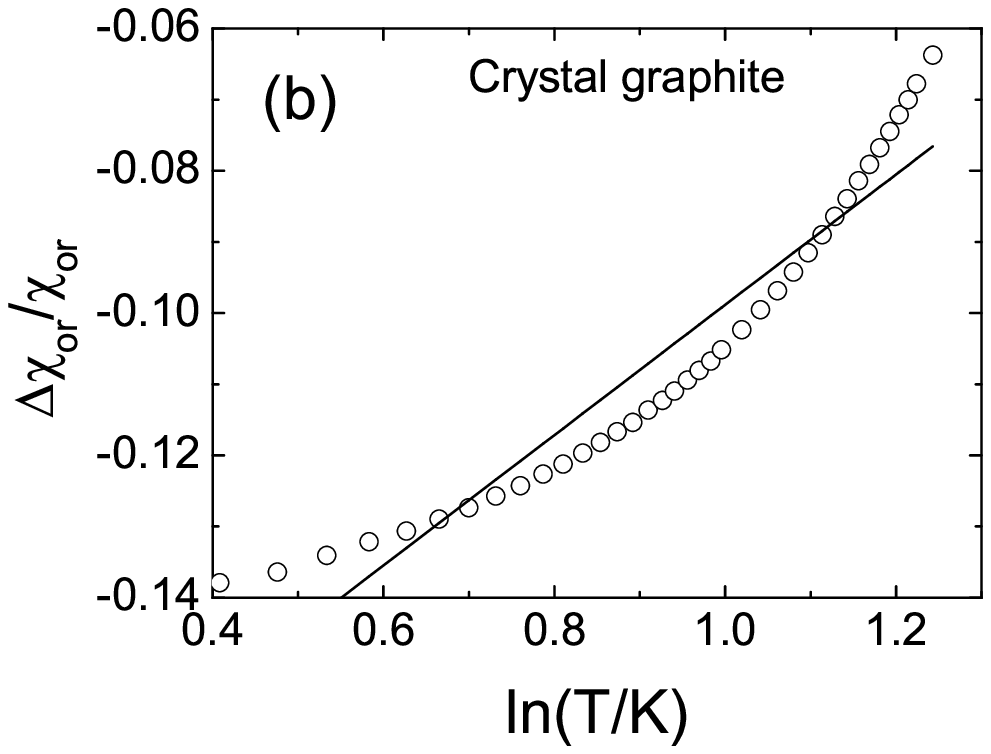}\cr
\includegraphics[height=1.7in]{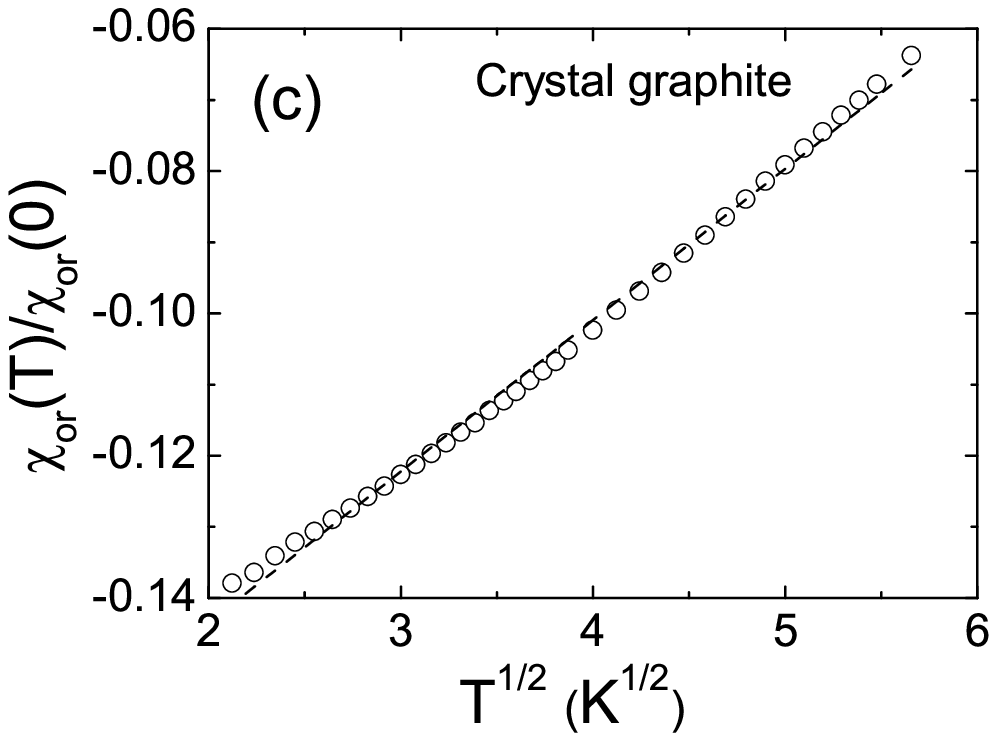}&
\includegraphics[height=1.7in]{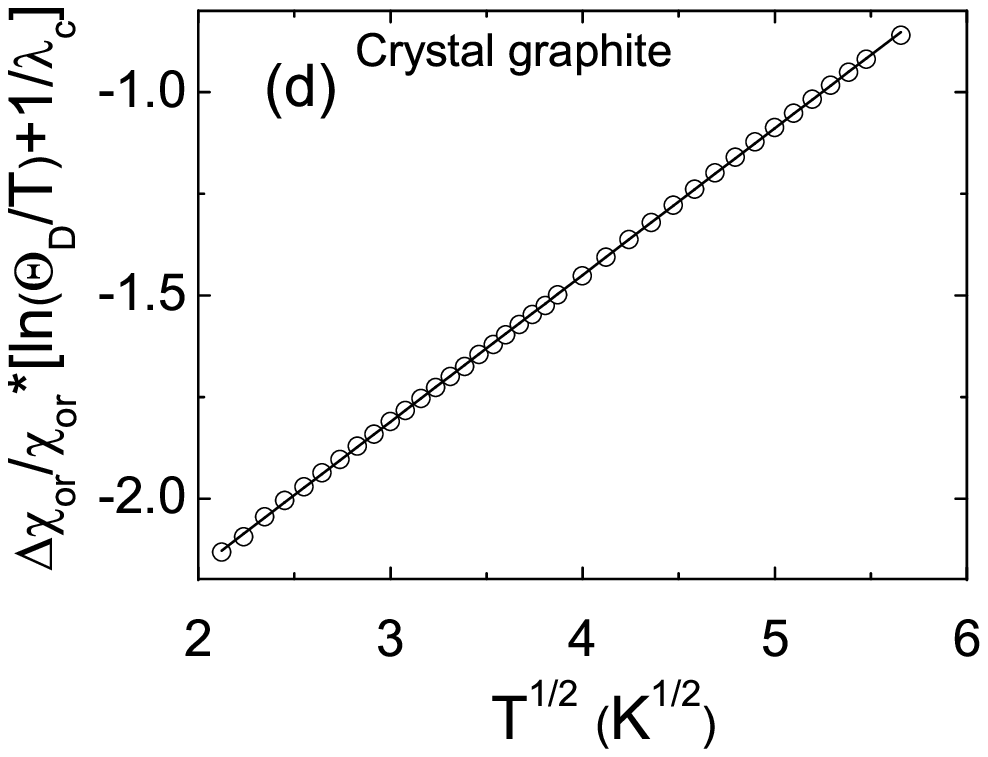}
\end{tabular}}
\caption[]{Temperature dependences of a magnetic susceptibility $\chi(T)$ for graphite measured in a magnetic field $Â$ = 0.01 T. Continuous lines show: the regular parts $\chi(T)$ (a); two-dimensional quantum corrections to $\chi(T)$ (b); three-dimensional quantum corrections to  $\chi(T)$ (c), (d). The solid lines are fits for ({\bf d}) by Eq. (3) in interval 4.5 - 45 K with parameters $\theta _D$ = 1000 K , $\lambda _c$ = 0.1.}\label{rofig1}
\end{figure}

The dependence of $\Delta \chi_{or} (T)/\chi _{or}(T)$ we
investigated for crystals of graphite (Fig. 3). However, only the
three-dimensional dependence $\Delta \chi_{or} (T)/\chi _{or}(T)
\approx T^{1/2}$ was found for graphite. 

Approximation of the
abnormal part of the magnetic susceptibility by theoretically
predicted functions has revealed three features:

I. A crossover from the two-dimensional quantum corrections to
$\chi(T)$ in fields $B$ = 0.01 T to the three-dimensional quantum
corrections in field $B$ = 5.5 T  is observed as up to bromination of arc-MWNTs, so
after bromination of its when the magnetic field increases. For
graphite, in all fields, the three-dimensional corrections to
$\chi(T)$ are observed. This is related to the fact that magnetic
length $L_B = 77\,${\AA} in fields of  $B =
5.5\,$T becomes comparable to the thickness of the graphene layers
$h_{MWNT}$ which form the MWNT. On the other hand, in fields of $B
= 0.01\,$T, $L_B = 1800\,{\mbox{\AA}}$ is
much longer than $h_{MWNT}$ but it is shorter than the length of a
tube $l_{MWNT}$ ($l_{MWNT}\approx 1 \mu)$ and is comparable to the
circumference of a nanotube. In graphite, the thickness of a
package of the graphene layers is always macroscopic and it
exceeds all other characteristic lengths.

II. The bromination of arc-MWNTs has led to an increase
of their conductance by one order of magnitude (from 500
$\Omega^{-1}$cm$^{-1}$ in arc-MWNTs before bromination up to 5000
$\Omega^{-1}$cm$^{-1}$ after bromination). However, the
relative correction to the magnetic susceptibility
$\Delta\chi(T)_{or}/\chi(T)_{or}$ which determines the value of $\lambda_c$,
remained constant. Thus, the bromination does not change the
electron-electron interaction constant $\lambda_c$ in the
arc-produced multiwalled carbon nanotubes.

III. The constant of electron-electron interaction $\lambda_c$ for arc-MWNTs 
before and after bromination has the 
magnitude about 0.2 \cite{Romanenko2} which is greater than that of graphite $\lambda_c
\approx 0.1$ \cite{Romanenko3}, i.e. the curvature of the graphene layers
in MWNTs leads to the increase of $\lambda_c$.

The temperature dependence of the electrical conductivity
$\sigma(T)$ of arc-MWNTs indicates the presence of quantum
corrections also (Fig. 4(a) and Fig. 5(a)). At low temperatures, the temperature dependence of
these quantum corrections is characteristic for the
two-dimensional case (Fig. 4(b) and Fig. 5(b)):

\begin{equation}
{\Delta\sigma(T)=\Delta\sigma_{WL}(T)+ \Delta\sigma_{IE}(T)}\,.
\end{equation}

Here $\Delta\sigma_{WL}(T)\approx {ln(L_{\varphi}/l_{el})}$ is the
correction associated with the quantum interference of
noninteracting electrons in  two-dimensional systems (WL)
\cite{Lee,Kawabata} while $\Delta\sigma_{IE}(T)\approx
ln(L_{IE}/l_{el})$ is the correction associated with the quantum
interference of interacting electrons (IE) in such systems
\cite{Lee,Altshuler}. The contribution of quantum corrections to the electrical
conductivity should be accompanied by corrections to
magnetoconductivity $\Delta \sigma(B)=1/ \rho(B)$ in low magnetic
fields \cite{Kawabata,Altshuler1}:

\begin{equation}
{\Delta\sigma(B)=\Delta\sigma_{WL}(B)+ \Delta\sigma_{IE}(B)}.
\end{equation}

Here $\Delta\sigma_{WL}(B)$ is the quantum correction to
magnetoconductance for noninteracting electrons;
$\Delta\sigma_{IE}(B)$ - the quantum correction to the
magnetoconductance for interacting electrons. Both corrections have the logarithmic asymptotic in high
magnetic fields ($\Delta\sigma_{WL}(B)\approx
ln(L_{\varphi}/L_B)$; $\Delta\sigma_{Int}(B)\approx
ln(L_{IE}/L_B)$ at $L_{\varphi}/l_B$; $L_{IE}/L_B>>1$), and the
quadratic asymptotic in low magnetic fields ($\Delta\sigma_{WL}(B)
\approx B^2$; $\Delta\sigma_{IE}(B) \approx B^2$ when
$L_{\varphi}/L_B$; $L_{IE}/L_B << 1$). The quantum corrections to
magnetoconductance become essential in low magnetic fields when
the magnetic length is $L_B<L _{\varphi}$.
In this case the phase of an electron is lost at
distances $\approx L_B$, and quantum corrections to conductance
are partially suppressed. This leads to positive
magnetoconductance (negative magnetoresistance).

It is difficult to divide the contribution of IE in the Cooper
channel (which is determined by the amplitude and sign of
$\lambda_c$), WL and IE in the diffusion channel. Field
dependences of magnetoresistance (Fig. 4(c)) in all intervals of
the measured fields (0 - 1 T)
 show negative magnetoresistance, related to
 WL. Similar dependences are observed in graphite
 (Fig. 4(d)). All three mechanisms give
 quantum corrections to temperature dependence of conductance
 $\sigma(T)$. In order to  observe the contribution of IE in the
 superconducting channel it is necessary to exclude the contribution
 of WL. We achieved this in catalytic carbon multiwalled nanotubes
 which were synthesized  by a technology which prevents
the formation  of other phases of carbon.

\begin{figure}
\footnotesize
\centerline{\begin{tabular}{@{}cc@{}}
\includegraphics[height=1.75in]{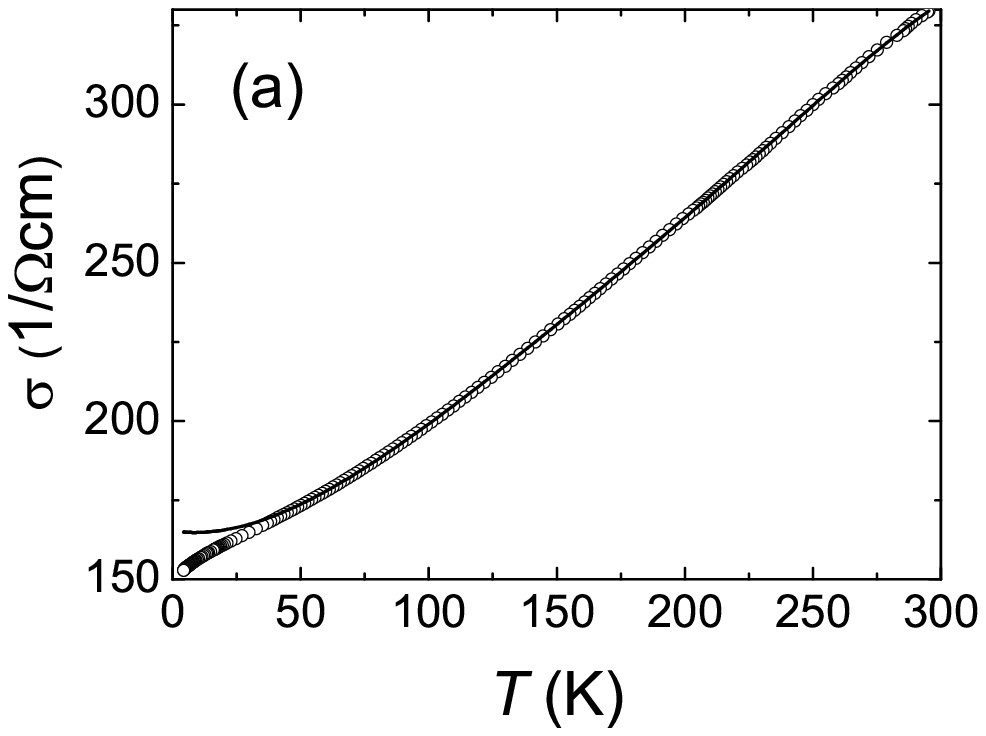}&
\includegraphics[height=1.75in]{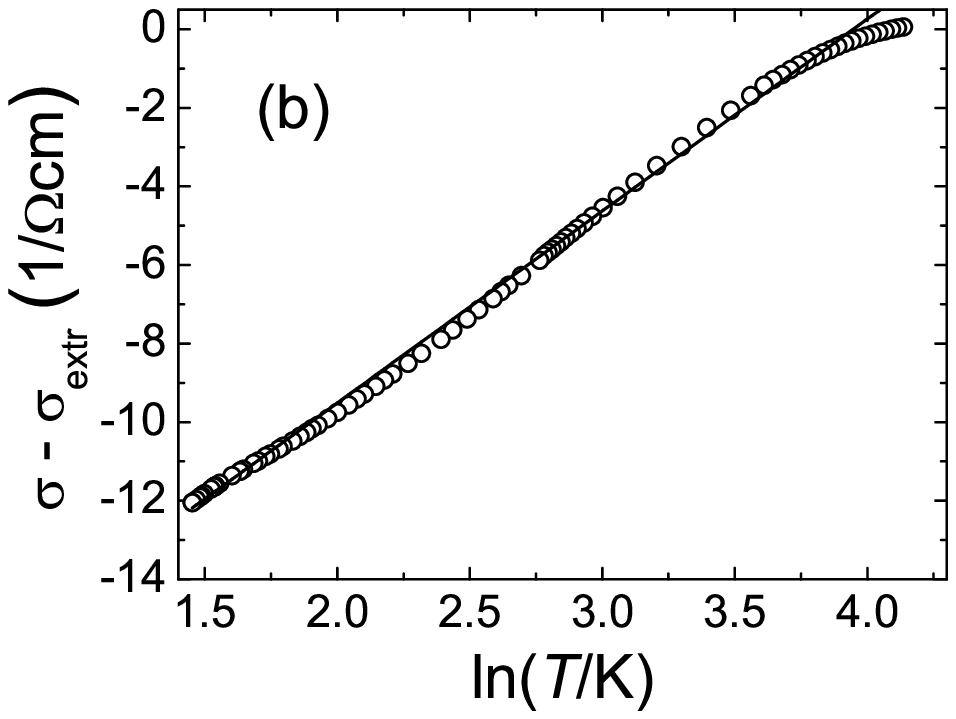}\\
\includegraphics[height=1.75in]{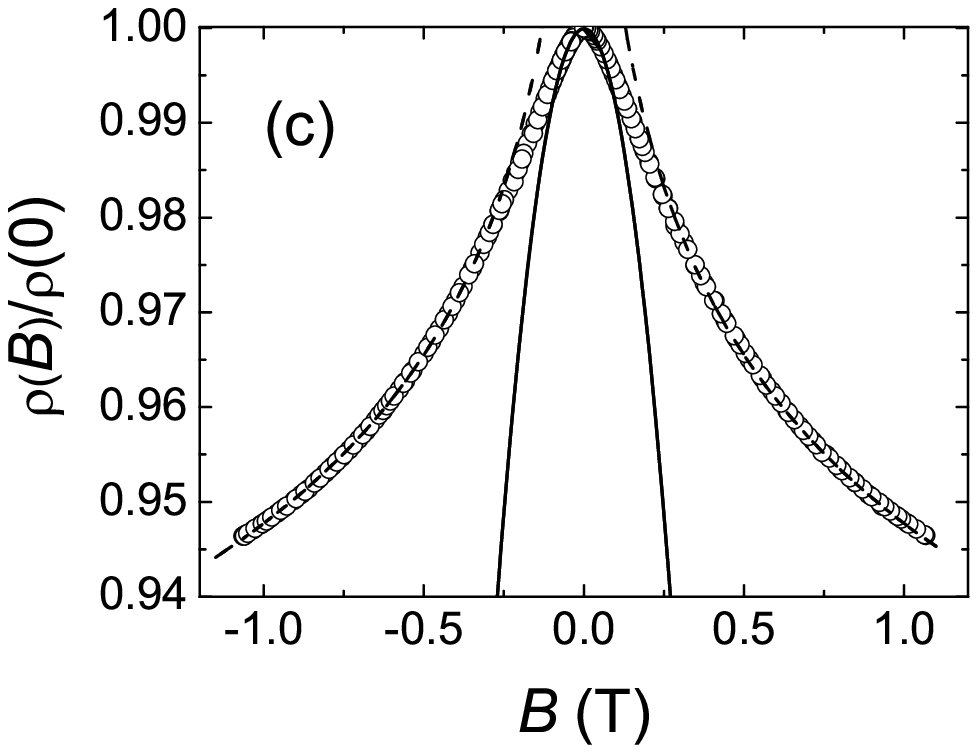}&
\includegraphics[height=1.8in]{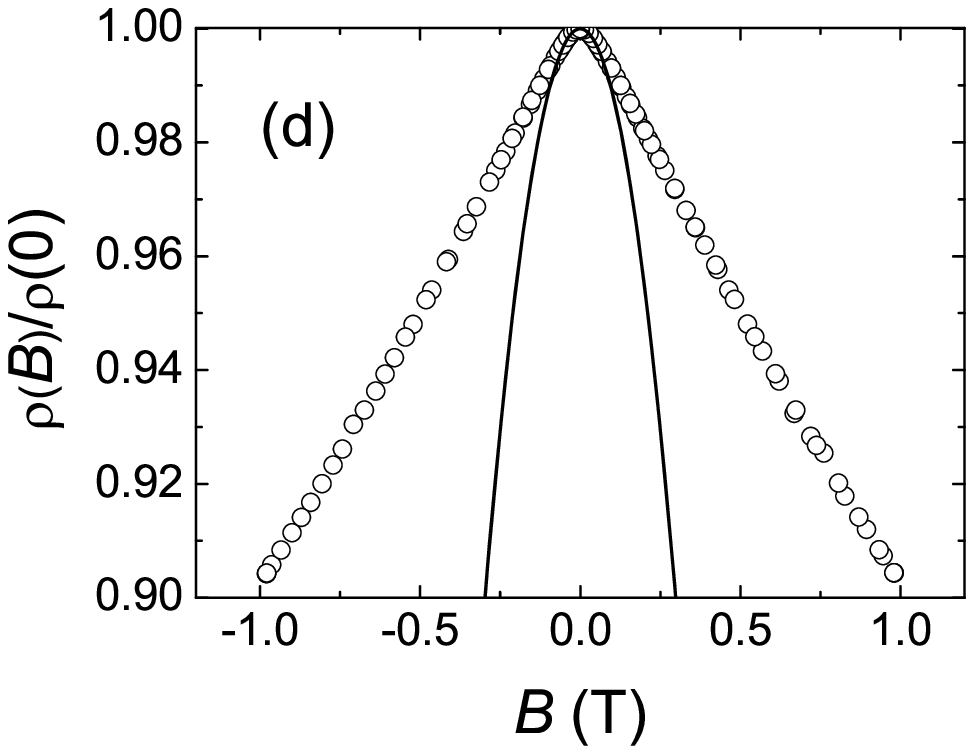}
\end{tabular}}
\caption[]{Data for arc-produced MWNTs (a, b, c) and for pyrolytic graphite (d). Temperature dependence of conductivity $\sigma(T)$ (a), anomaly part of conductivity $\Delta\sigma(T) = \sigma(T) - \sigma(T)_{ext}$ (b), and the relative magnetoconductivity  $\sigma(B)/\sigma(0)$ from magnetic field $B$ measured at $T$ = 4.2 K (c, d). Continuous lines show: $\sigma(T)_{ext}$ receiving by extrapolation of approximation curve from $T \geq 50$ K to $T \leq 50$ K (a); two-dimensional quantum corrections to $\sigma(T)$ (b), asymptotic of quadratic approximation $\sigma(B)/\sigma(0)\approx B^2$ at $B \leq 0.05$ T to $B$ up to 0.3 T (c, d). Dashed lines on (c) show the logarithmic asymptotic $\sigma(B)/\sigma(0) \approx ln(T)$ from high field to low field.}\label{rofig2}
\end{figure}

\subsection{Catalytic multiwalled carbon nanotubes}
There are always paramagnetic impurities present in c-MWNTs  because of the
ferromagnetic metals used as catalytic agents. It is not possible
to use the magnetic susceptibility in order to obtain information
on $\Delta \chi_{or} (T)$ of c-MWNT, because the contribution of
paramagnetic impurities dominates at low temperatures. We carried
out an analysis of the conductivity data for which the
contribution of paramagnetic impurities is negligible. In figure
5(c) the dependence of the relative magnetoconductivity
$\rho(B)/\rho(0)$ on the magnetic field $B$ for c-MWNTs is shown.

\begin{figure}
\footnotesize
\centerline{\begin{tabular}{@{}cc@{}}
\includegraphics[height=1.75in]{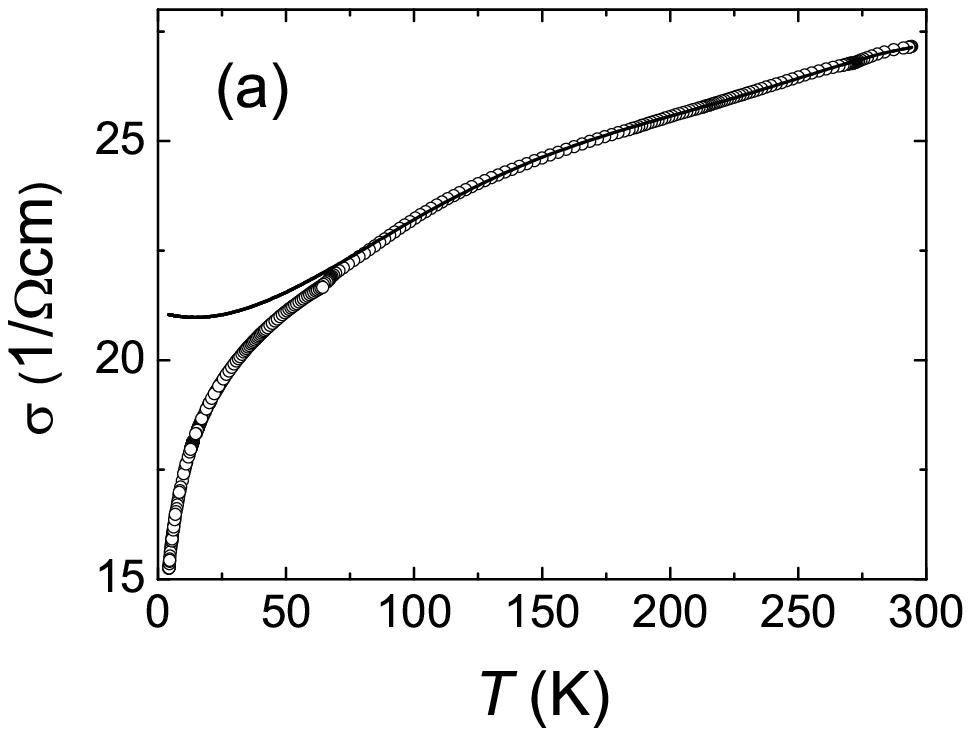}&
\includegraphics[height=1.75in]{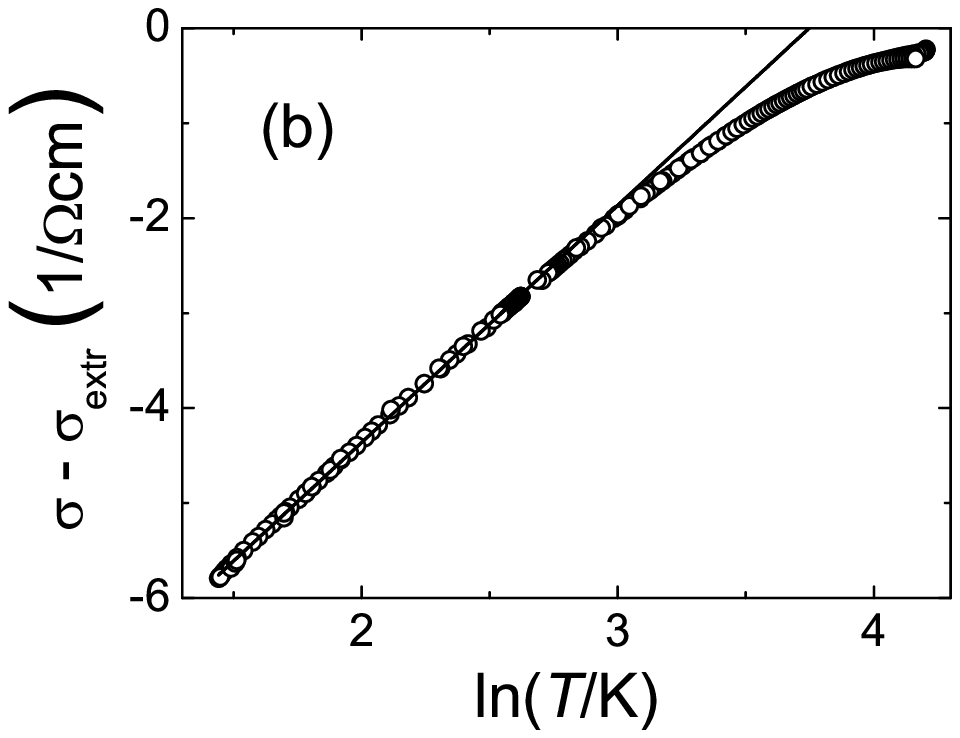}\\
\includegraphics[height=1.75in]{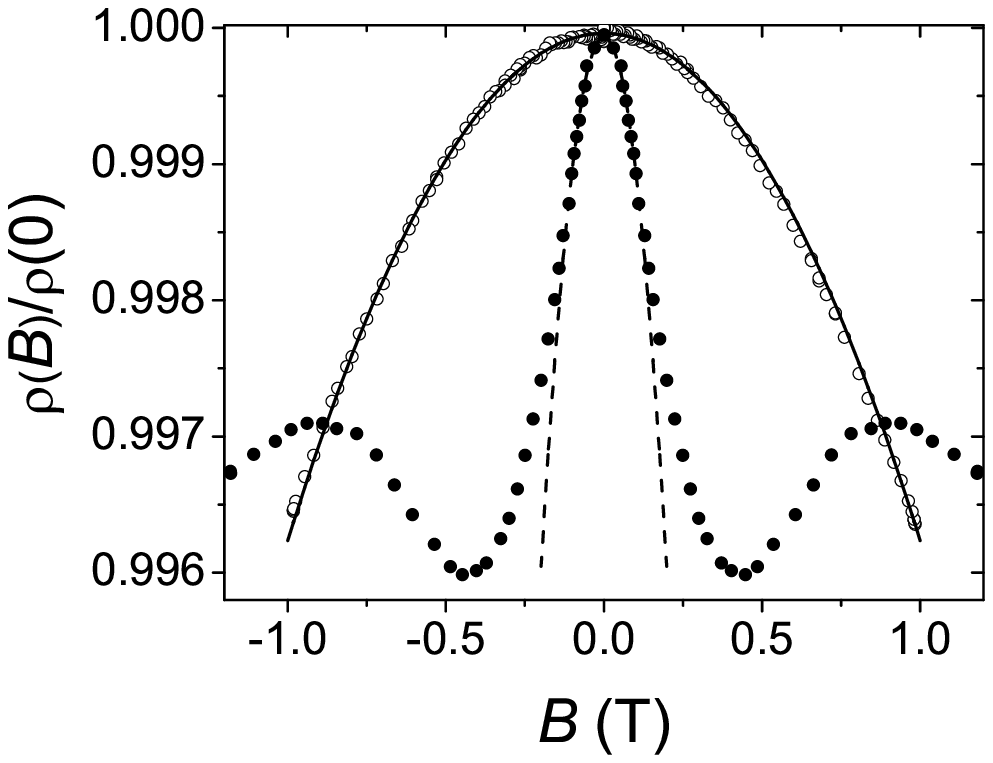}&
\includegraphics[height=1.8in]{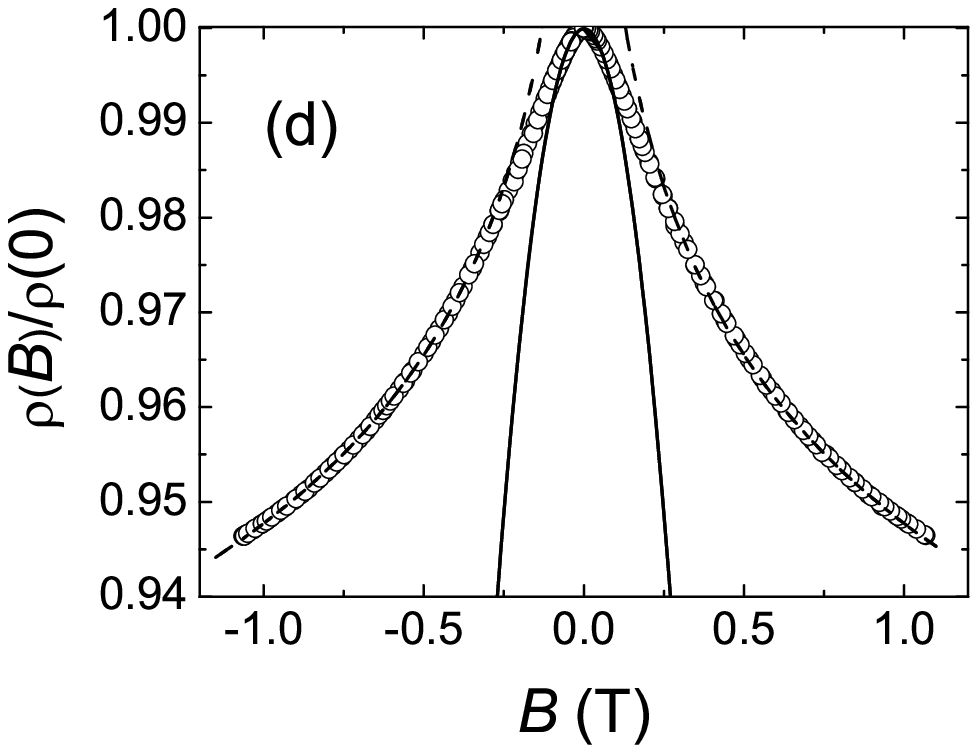}
\end{tabular}}
\caption[]{Data for: $\circ$ - catalytic MWNTs without impurities of another forms of carbon (a, b, c); $\bullet$ - MWNTs prepared by a usual catalytic method(c), and $\Delta$ - soot (d). Temperature dependence of conductivity $\sigma(T)$ (a), anomaly part of conductivity $\Delta\sigma(T) = \sigma(T) - \sigma(T)_{ext}$ (b), and the relative magnetoconductivity $\sigma(B)/\sigma(0)$ from magnetic field $B$ measured at $T$ = 4.2 K (c, d). Continuous lines show: $\sigma(T)_{ext}$ receiving by extrapolation of approximation curve from $T \geq 50$ K to $T \leq 50$ K (a), two-dimensional quantum corrections to $\sigma(T)$ (b), asymptotic of quadratic approximation $\sigma(B)/\sigma(0)\approx B^2$ from $B \leq 0.05$ T to $B$ up to 0.3 T for  (curve $\bullet$ on figure c) and soot (d). Dashed line: on (c) - show the quadratic approximation $\sigma(B)/\sigma(0) \approx B^2$, on (d) - show the logarithmic asymptotic $\sigma(B)/\sigma(0) \approx ln(T)$ from high field to low field.}\label{rofig3}
\end{figure}

The closed circles - c-MWNTs prepared by a usual catalytic method
\cite{Kudashov}; the open circles - with use of the special
procedure \cite{Couteau} which allows to prepare c-MWNTs
practically without inclusions of the another form of carbon.
Comparing these curves it is possible to see, that in the refined
samples in the low magnetic fields (at $B \leq 0.2$ T) the contribution to the
negative magnetoconductivity, related to WL, we not observe.
$\sigma(B)/\sigma(0)$ for refined c-MWNTs is described by
quadratic dependence $\sigma(B)/\sigma(0) \approx B^2$ for all
values of field $B$. $\sigma(B)/\sigma(0)$ for c-MWNT prepared by
a usual method are described by quadratic dependence only in low
magnetic fields ($B < 0.1$ T). In figure 5(d) the dependence of
the relative magnetoconductivity $\sigma(B)/\sigma(0)$ on the
magnetic field $B$ is shown for soot. As can be seen from figure 5
the curves for soot and c-MWNTs prepared by a usual catalytic
method are very similar at low fields. We suggest that this fact
is connected to the presence of impurities of soot in c-MWNTs
prepared by a usual catalytic method.
Investigation of $\sigma(T)$ of the c-MWNTs shows the presence of
quantum corrections $\sigma(T)\approx ln(T)$ and suggests the
two-dimensional character of these corrections \cite{Kawabata,
Lee, Altshuler}. The magnitude of the magnetic field which
suppresses the temperature correction ($\delta\sigma$(4.2
K)/$\sigma$(4.2 K) $\approx 2.7\%$) is estimated as $B \approx
8.5$ T. This corresponds to a quite reasonable magnitude of
magnetic length $L_B \approx ~ 60\,${\AA}.
Negative magnetoconductivity for IE is also in agreement with the
conclusion that $\lambda_c
> 0$. However this result has been already obtained from the
resistance data.

\section{Conclusion}
Analysis of the anomalous part of the  magnetic susceptibility
$\chi$ at temperatures below 50 K has allowed us to estimate the
electron-electron interaction constant, $\lambda_c$, in the
arc-produced MWNTs ($\lambda_c \approx 0.2$) and in graphite
($\lambda_c \approx 0.1$). We found that $\lambda_c$ does not
change in arc-produced MWNTs when the concentration of current
carriers is modified by bromination. The analysis of the anomalous
part of the conductance and the positive magnetoconductivity
demonstrate a dominating of contribution of weak localization effects.
In pure catalytic MWNTs we found positive magnetoconductivity
related only to interaction effects which points to the positive
sign of $\lambda_c$ in these nanotubes. Thus, the analysis of
temperature and field dependences of the magnetic susceptibility,
conductivity and magnetoconductivity allows us to estimate the
electron-electron interaction constant, $\lambda_c$, and to
determine the effective dimensionality of the current carriers in
inhomogeneous systems. On the base of our results we can conclude that the curvature of graphine layers in carbon
nanostructures is responsible for the change of constant electron-electron interaction $\lambda_c$.

\acknowledgements
The work was supported by Russian Foundation of Basic Research (Grants
No: 05-03-32901, 06-02-16005); Russian Ministry of
Education and Sciences (Grant PH$\Pi$.2.1.1.1604); Joint Grant CRDF, and Russian
Ministry of Education and Sciences (NO-008-X1).

%\nocite{*} % add all entries from sample.bib

\bibliographystyle{nato}
\bibliography{natoman}

\begin{thebibliography}{}



\bibitem[\protect\citeauthoryear{Al'tshuler et~al.}{1983}]{Altshuler}
Al'tshuler, B. L., Aronov, A. G., and Zyuzin A. Yu. (1983)
Thermodynamic properties of disordered conductors, {\em
Sov. Phys. JETP} {\bf 57}, 889--895.

\bibitem[\protect\citeauthoryear{Al'tshuler et~al.}{1981}]{Altshuler1}

Al'tshuler, B. L., Aronov, A.G., Larkin, A.I., and Khmel'nitski,
D.E. (1981) Anomalous magnetoresistance in semiconductors, {\em
Sov. Phys. JETP} {\bf 54}, 411--419.

\bibitem[\protect\citeauthoryear{Baxendale et~al.}{1997}]{Baxendale}
Baxendale, M., Mordkovich, V.Z., Yoshimura, S., and Chang, R.P.H. (1997) Magnetotransport in bundles of intercalated carbon nanotubes, {\em Phys. Rev. B} {\bf 56}, 2161--2165.

\bibitem[\protect\citeauthoryear{Couteau et~al.}{2003}]{Couteau}
Couteau, E., Hernadi, K., Seo, J.W., Thien-Nga, L., Miko, C., Gaal, R., Forro, L. (2003) CVD synthesis of high-purity multiwalled carbon nanotubes using CaCO3 catalyst support for large-scale production, {\em Chem. Phys. Lett.} {\bf 378}, 9-17.

\bibitem[\protect\citeauthoryear{Gonzalez et~al.}{2001}]{Gonzalez}
Gonzalez, J., Guinea, F., and Vozmediano, M. A. H. (2001) Electron-electron interactions in graphene sheets, {\em Phys. Rev. B} {\bf 63}, 134421-1 --134421-8.

\bibitem[\protect\citeauthoryear{Kawabata}{1980}]{Kawabata}
Kawabata A. (1980) Theory of negative magnetoresistance in three-dimensional systems, {\em Solid State Commun.} {\bf 34}, 431--432.

\bibitem[\protect\citeauthoryear{Kociak et~al.}{2001}]{kociak}
Kociak, M., Kasumov, A.Yu., Guron, S., Reulet, B., Khodos, I. I., Gorbatov, Yu. B., Volkov, V. T., Vaccarini, L., and Bouchiat, H. (2001) Superconductivity in Ropes of Single-Walled Carbon Nanotubes, {\em Phys. Rev. Lett.} {\bf 86}, 2416--2419.

\bibitem[\protect\citeauthoryear{Kopelevich et~al.}{2000}]{Kopelevich}
Kopelevich, Y., Esquinazi, P., Torres,J. H. S., and Moehlecke, S. (2000) Ferromagnetic- and Superconducting-Like Behavior of Graphite, {\em Journal of Low Temperature Physics} {\bf 119} 5--6.

\bibitem[\protect\citeauthoryear{Kotosonov et~al.}{1997}] {Kotosonov} 
Kotosonov, A. S., Kuvshinnikov, S. V. (1997) 
Diamagnetism of some quasi-two-dimensional graphites and multiwall carbon nanotubes,
{\em Phis. Lett. A\/} {\bf 229} 377--380.

\bibitem[\protect\citeauthoryear{Kotosonov et~al.}{1997}] {Kotosonov} 
97) Diamagnetism of some quasi-two-dimensional graphites and multiwall carbon nanotubes, {\em Phys. Lett. A} {\bf 229}, 377--380

\bibitem[\protect\citeauthoryear{Kudashov et~al.}{2002}]{Kudashov}
Kudashov, A. G., Okotrub, A. V., Yudanov, N. F., Romanenko, A. I., Bulusheva, L. G., Abrosimov, O. G., Chuvilin, A. L., Pazhetov, E. M., Boronin, A. I. (2002) Gas-phase synthesis of nitrogen-containing carbon nanotubes and their electronic properties, {\em Physics of Solid State} {\bf 44} 652--655.

\bibitem[\protect\citeauthoryear{Kudashov et~al.}{2004}]{Kudashov4}
Kudashov, A. G., Abrosimov, O. G., Gorbachev, R. G., Okotrub, A. V., Yudanova  L. I., Chuvilin, A. L., Romanenko, A. I. (2004) Comparison of structure and conductivity of multiwall carbon nanotubes obtained over Ni and Ni/Fe catalysts, {\em Fullerenes, Nanotubes, and Carbon Nanostructures} {\bf 12}, 93--97.

\bibitem[\protect\citeauthoryear{Lee and Ramakrishnan}{1985}]{Lee}
Lee P A, Ramakrishnan T. V. (1985) Disordered electronic systems, {\em Rev. Modern Phys.} {\bf 57}, 287--337.

\bibitem[\protect\citeauthoryear{Mott}{1979}]{Mott}
Mott, N. F. (1979) Electron Processes in Noncrystalline Materrials, {\em Oxford, Clarendon Press}, 350 p.

\bibitem[\protect\citeauthoryear{Okotrub et~al.}{2001}]{Okotrub}
Okotrub, A.V., Bulusheva, L.G., Romanenko, A.I., Chuvilin, A.L., Rudina, N.A., Shubin, Y.V., Yudanov, N.F., Gusel'nikov, A.V. (2001) Anisotropic properties of carbonaceous material produced in arc discharge, {\em Appl. Phys. A} {\bf 71} 481-486.

\bibitem[\protect\citeauthoryear{Okotrub et~al.}{2001}]{Okotrub01}
Okotrub, A. V., Bulusheva, L. G., Romanenko, A. I., Kuznetsov, V.L., Butenko, Yu.V., Dong, C., Ni, Y., Heggic, M.I. (2001) Probing the electronic state of onion-like carbon. in Electronic Properties of Molecular Nanostructures (AIP Conference Proceedings, New York, 2001) 591 p.349--352.

\bibitem[\protect\citeauthoryear{Romanenko et~al., PSS}{2002}]{Romanenko}
Romanenko, A. I., Anikeeva, O. B., Okotrub, A. V., Bulusheva, L. G., Yudanov, N. F., Dong, C., and Ni, Y. (2002) Transport and Magnetic Properties of Multiwall Carbon Nanotubes before and after Bromination, {\em Physics of Solid State} {\bf 44}, 659--662.

\bibitem[\protect\citeauthoryear{Romanenko et~al., SSC}{2002}]{Romanenko2}
Romanenko, A. I., Okotrub, A. V., Anikeeva, O. B., Bulusheva, L. G., Yudanov, N. F., Dong, C., Ni, Y. (2002) Electron-electron interaction in multiwall carbon nanotubes, {\em Solid State Commun.} {\bf 121}, 149--153.

\bibitem[\protect\citeauthoryear{Romanenko et~al., Pit}{2002}]{Romanenko002}
Romanenko, A. I., Anikeeva, O. B., Okotrub, A. V., Kuznetsov, V.L., Butenko, Yu.V., Chuvilin, A.L., Dong, C., Ni, Y. (2002) Diamond nanocomposites and onion-like carbon in Nanophase and nanocomposite materials, vol. 703, Eds. S. Komarneni, J.-I. Matsushita, G.Q. Lu, J.C. Parker, R.A. Vaia, Material Research Society, (Pittsburgh 2002) p. 259--264 Temperature dependence of electroresistivity, negative and positive magnetoresistivity of graphite.

\bibitem[\protect\citeauthoryear{Romanenko et~al.}{2003}]{Romanenko3}
Romanenko, A. I., Okotrub, A. V., Bulusheva, L. G., Anikeeva, O. B., Yudanov, N. F., Dong, C., Ni, Y. (2003) Impossibility of superconducting state in multiwall carbon nanotubes and single crystal graphite, {\em Physica C} {388-389}, 622--623.

\bibitem[\protect\citeauthoryear{Romanenko et~al.}{2004}]{Romanenko4}
Romanenko, Romanenko, Anikeeva, O. B., Zhmurikov, E. I., Gubin, K. V., Logachev, P.V., Dong, C., Ni, Y. (2004)  Influence of the structural defects on the electrophysical and magnetic properties of carbon nanostructures, {\em in  The Progresses In Function Materials (11th APAM Conference proceedings}, Ningbo, P. R. China, 2004) 59--61.

\bibitem[\protect\citeauthoryear{Takesue et~al.}{2006}]{takesue}
Takesue, I., Haruyama, J., Kobayashi, N., Chiashi, S., Maruyama,  S., Sugai, T., Shinohara, H., (2006) Superconductivity in Entirely End-Bonded Multiwalled Carbon Nanotubes, {\em Phys. Rev. Lett.} {\bf 96}, 057001-1--057001-4.

\bibitem[\protect\citeauthoryear{Tang et~al.}{2003}]{tang}
Tang, Z.K., Zhang, L.Y., Wang, N., Zhang, X.X., Wang, J.N., Li, G.D., Li, Z.M., Wen, G.H., Chan, C.T., Sheng, P. (2003) Ultra-small single-walled carbon nanotubes and their superconductivity properties, {\em Synthetic metals} {\bf 133-134}, 689--693


\end{thebibliography}

% content of natoman.bbl:
%

% The endnotes section will be placed here.

%\theendnotes
%\newpage

%\printindex
\end{document}